\documentstyle[prd,aps,psfig]{revtex}

\bibstyle{unsrt}

\tighten
\begin{document}
\draft

\preprint{SUSX-TH-98-015}
\twocolumn[\hsize\textwidth\columnwidth\hsize\csname 
@twocolumnfalse\endcsname

\title{The length distribution of vortex strings  
in $U(1)$ equilibrium scalar field theory}
\author{Nuno D. Antunes$^1$ and Lu\'{\i}s M. A.  Bettencourt$^2$}
\address{$^1$School of Mathematical and Physical Sciences,
University of Sussex, Brighton BN1 9QH, U.K.} 
\address{$^2$T6/T11 Theoretical Division MS B288, 
Los Alamos National Laboratory, Los Alamos NM 87545}
\date{\today}
\maketitle

\begin{abstract}
We characterize the statistical length distribution of vortex strings
by studying the non-perturbative thermodynamics of scalar 
$U(1)$ field theory in 3D. All parameters characterizing the
length distributions exhibit critical behavior at $T_c$ and we measure
their associated critical exponents.   
Moreover we show that vortex strings are generally 
self-seeking at finite temperature and that there exists a natural 
separation between long strings and small loops based on 
their spatial correlation behavior.
\end{abstract}

\pacs{PACS Numbers : 05.70.Fh, 11.27.+d, 98.80.Cq \hfill   
SUSX-TH-98-015 }

\vskip2pc]
 
The accurate knowledge of the thermodynamic density and length 
distribution of a population of topological defects is a 
fundamental ingredient for scenarios of defect formation and 
evolution in the early Universe and in the laboratory.
Although a phenomenological model exists, 
due to Kibble \cite{Kibble} and Zurek \cite{Zurek}, 
for predicting defect densities
it does not provide us  with a microscopic understanding of the 
dynamics of defect populations, their length distribution 
and  correlations. 
These aspects of the defect network at the time of formation
will have a crucial impact on the subsequent network evolution.
  
In the past, several attempts have been made at characterizing 
the statistics of defect networks. 
Analytical studies of cosmic  string distributions at finite temperature
were based on the assumption that strings form a gas of 
non-interacting random walks \cite{EdRay}. The generalization
to realistic interactions is problematic and yields
only qualitative predictions.
Numerical support for the free random walk picture was 
found from Vachaspati-Vilenkin type algorithms \cite{VV}.
In these simulations phases of an underlying field are thrown down 
at random on sites of a lattice and vortex lines are formed by joining points 
around which the phases wind about the circle following the shortest 
path. This yields a (lattice-type dependent) 
fraction of long string of $67-80 \%$ and a length distribution 
that is Brownian for long enough strings. 
However these studies correspond to the {\it infinite} temperature
limit of the underlying field theory \cite{Us}, 
where it becomes trivially free, and therefore have no direct 
predictive power at criticality, when defects are formed.   

Recently, developments in computer power made 
possible the first studies of defect network properties directly 
from the nonperturbative thermodynamics of field theories. These
focused their attention on defect densities. 
In this letter we proceed to characterize in detail
the  equilibrium statistics of a network of vortex strings in 
scalar $U(1)$ field theory in 3D. We measure several universal critical 
exponents associated with the string length distribution and 
construct a detailed picture of the critical behavior of the theory 
in terms of vortex strings.

The adoption of a dual perspective, {\it i.e.} of the description 
of the mechanisms underlying criticality in terms of non-linear
solutions of the theory was found to be the key to understanding 
the Kosterlitz-Thouless \cite{KT} transition in 2D, which proceeds by 
vortex-antivortex pair unbinding.
Generally, given a dual model, critical behavior can be described 
equivalently in either the fundamental or the dual 
picture, the choice between either point of view being made
based on computational advantage. However, the construction of 
exact dual theories proves very difficult in $D \geq 3$, except 
when supersymmetry is unbroken.
The measurements of the string network universal critical exponents 
presented below provide us with direct constraints on dual 
field theories to $U(1)$ $\lambda \phi^4$ in 3D \cite{Adriaan}.

Although $U(1)$  $\lambda \phi^4$ is the simplest  
field theory containing vortex lines the model is of 
direct relevance to the thermodynamic 
description of liquid $^4He$ and of strong type II 
superconductors (such as most new 
high-$T_c$ materials) in the absence of an external field. It also 
embodies applications to cosmology, eg.,  the phase transition 
associated with the spontaneous breakdown of Peccei-Quinn symmetry 
and the formation of axionic strings \cite{VE}.

The methods used to drive the system to canonical equilibrium 
are familiar from stochastic field quantization.
A more detailed description can be found elsewhere \cite{Us}.
The equations of motion for the fields $\phi_i$ are  
\begin{eqnarray}
\left( \partial_{t}^2 -\nabla^2 \right) \phi_i - m^2 \phi_i
+ \lambda \phi_i \left( \phi_i^2 + \phi_j^2 \right) 
+  \eta \dot{\phi_i} = \Gamma_i 
\label{e1}
\end{eqnarray}
where $i,j \in \{1,2\}$  and $i \neq j$ in Eq.~(\ref{e1}). 
$\Gamma_i(x)$ is the  Gaussian noise characterized by 
\begin{eqnarray}
\langle \Gamma_i(x) \rangle = 0, 
\qquad  \langle \Gamma_i(x) \Gamma_j(x')\rangle 
= {2 \eta \over \beta} \delta_{ij} \delta (x-x'),
\label{e2}
\end{eqnarray}
where $x,x'$ denote space-time coordinates. Eqs.~(\ref{e1}-\ref{e2}) 
ensure that the fluctuation-dissipation theorem applies 
in equilibrium, which will result for large times at temperature $T=1/\beta$.
We use a computational domain of size $N^3=100^3$, lattice spacing 
$\Delta x=0.5 $ and impose periodic boundary conditions in space.
At criticality all universal exponents are independent of $\Delta x$.
We assume ergodicity of the evolution after equilibrium was reached. 
Using this fact we construct the string network from the phase of the 
complex field $\Phi=\phi_1 + i \phi_2$ , 
at regular time intervals. In all measurements presented
below a number of samples between 400 and 1200 was used. This permits us 
studies of the string network with unprecedented high statistics.
The inverse critical temperature $\beta_c$, measured from field 
correlators, is $\beta_c = 1.91 \pm 0.01$
\cite{Us}. 

\begin{figure}
\centerline{\psfig{file=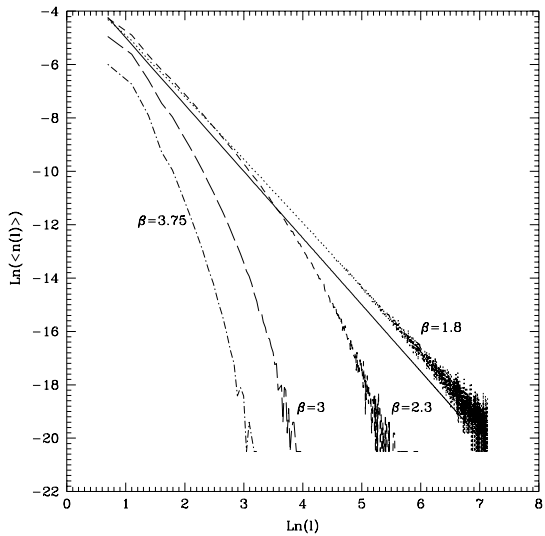,width=3in}} 
\caption{The loop length distribution for several $\beta$. 
An arbitrary $l^{-{5 \over 2}}$ distribution (solid) is plotted 
for comparison.} 
\label{fig1}
\end{figure}

We separate strings as has become usual in cosmology
into two distinct populations, one of loops, 
vortex lines shorter than $\sim N^2$ and another 
of long strings longer than any loop. 
Note that there is {\it a priori} no  reason for this 
separation or for the definition of the boundary between the two classes. 
First we deal with the statistics of loops.
Characteristic loop distributions above and below the critical 
temperature are shown in Fig.~\ref{fig1}.  
   
In order to characterize the loop length distribution $n(l)$ quantitatively we 
fit it for each temperature to an expression of the form:
\begin{equation}
 \langle n(l) \rangle  
= A~l^{- \gamma}~\exp( - \beta \sigma_{\rm eff} l),
\label{e3}
\end{equation}
where the average is over the canonical ensemble.
$\sigma_{\rm eff}$ is the $T$ dependent effective string tension 
and has dimensions of mass per unit length. The temperature at which it 
vanishes defines the so-called Hagedorn 
temperature $T_H$, the point at which long string appears in the system. 
We present our estimate of errors, mostly resulting from 
performing multiparameter fits, in the final measurements of universal 
quantities only. Statistical errors are generally very small. 
$\gamma$ is related to the string interactions. 
For Brownian (i.e. non-interacting) strings $\gamma=5/2$, whereas  
$\gamma>5/2$ or $\gamma<5/2$ for self-avoiding or self-seeking walks, 
respectively.
The loop distributions measured in equilibrium  are in fact very well fit by 
Eq.(\ref{e3}), with 
$A(\beta)$, $\gamma(\beta)$ and $\sigma_{\rm eff}(\beta)$ determined
by $\chi^2$ minimization. 
These three quantities undergo radical behavior changes near the $T_c$ 
lending support to a recent claim \cite{Us} 
that the critical temperature for the fields, $T_c$  
and the Hagedorn temperature for the strings, $T_H$ 
coincide in the infinite volume limit.

\begin{figure}
\centerline{\psfig{file=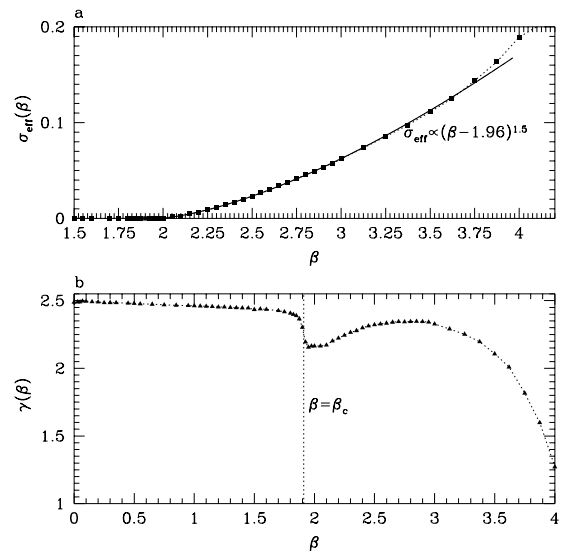,width=3in}}
\caption{a) $\sigma_{\rm eff}(\beta)$ and b) $\gamma(\beta)$}
\label{fig2}
\end{figure}

Fig.~\ref{fig2} shows the temperature dependence of $\gamma$ and
$\sigma_{\rm eff}$. We observe that as the critical point is approached from 
below $\sigma_{\rm eff}$ vanishes. A fit to $\sigma_{\rm eff}$
in this regime gives
\begin{eqnarray}
&& \sigma_{\rm eff} (\beta) \propto (\beta-\beta_H)^{\nu_{st}}\nonumber
\end{eqnarray}
with $\beta_H=1.96 \pm 0.01$, $\nu_{st}=1.50 \pm 0.01$ and 
where $\beta_H=1/T_H$. Note that $\beta_H$ 
does not coincide exactly with $\beta_c$.
This discrepancy was observed previously in \cite{Us}, where it was shown that 
as the computational domain is increased the temperature at which 
long strings first appear $\beta_H$ migrates to $\beta_c$.  
In addition the relative precision of the measurement of 
$\sigma_{\rm eff}$ close to $\beta_c$ is poor since 
it requires a large number of long loops, lying on the boundary 
of the long string/loop separation, eg., for $\beta=1.975$, we 
have that $l \sim (\beta \sigma_{\rm eff})^{-1} = 0.44\times 10^4\sim N^2$.
Below a more accurate measurement of $\beta_H$ will be obtained from the 
behavior of the length distribution of long strings.  

Also very interesting is the value of $\nu_{\rm st}$. This universal 
exponent for $\sigma_{\rm eff}$ in the global $U(1)$ theory, corresponds
to the exponent $\nu_{\rm dual}=\nu_{\rm st}/2$, 
associated with the correlation length of its dual field theory 
\cite{Adriaan}. 
We believe this measurement provides a first direct determination
leading to the classification of the universality class to which the dual
formulation belongs. 
Note that $\nu_{\rm dual}=0.75$ differs clearly from the mean-field 
prediction $\nu_{\rm MF}=0.5$, for any $O(N)$, $N \geq 1$, 
scalar field theory. 
Remarkably $\nu_{\rm dual}$ coincides with the {\it exact} 
value for the critical exponent of an ensemble of interacting 
polymers in 2D \cite{Zinn}.

As for $\gamma$, it is especially interesting to note that $\gamma<5/2$ 
generally at any finite temperature, see Fig.~\ref{fig2}.
The discrepancy between 
our observations and the Brownian result cannot be attributed to small
computational domain effects. The computation of the Brownian distribution 
in a finite volume generates exponential corrections \cite{ERA}
to Eq.~(\ref{e3}), which fail to describe the changes in the 
distributions correctly and are moreover $\beta$ independent. 
Above $T_c$, $\gamma(\beta) \simeq 5/2 - 0.0385~\beta$ and 
$\gamma(\beta_c)= 2.23 \pm 0.04$, \cite{Williams}.

\begin{figure}
\centerline{\psfig{file=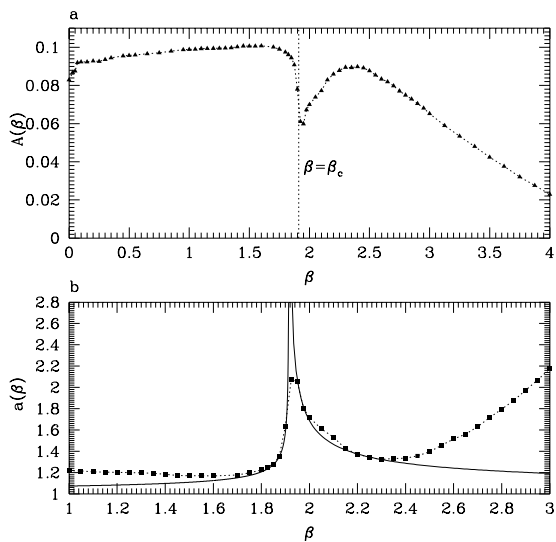,width=3in}}
\caption{a) The amplitude of the loop distribution $A(\beta)$ and 
b) the average step length $a(\beta)$ and the corresponding best fits
Eqs.~(\ref{e5}) at criticality. }
\label{fig3}
\end{figure}

Fig.~\ref{fig3} shows the $\beta$ dependence of $A$ and of 
the average {\it walk step size} $a(\beta)$,
defined in terms of $A(\beta)$ and $\gamma(\beta)$ by
\begin{eqnarray}
A = \left({3 \over 2 \pi} \right)^{3/2} a^{\gamma -3}.
\label{e4}
\end{eqnarray}
The numerical prefactor in Eq.~(\ref{e4}) is obtained from 
the computation of $\langle n(l) \rangle$ in the 
free random walk case \cite{ERA}.
We observe that the amplitude of the distribution $A(\beta)$ 
varies drastically with temperature in the vicinity 
of $T_c$. 
$A(\beta)$ alone determines the $\beta$-dependent 
probability of finding the smallest loops.
In the vicinity of the critical point 
$A$ suffers a dramatic drop until $\beta_c$ is hit, followed
by a sudden increase. At $\beta=\beta_c$ $dA/d\beta$ is discontinuous.
The sharp drop in $A$ signifies that $a(\beta)$ has grown and with it 
the possibility for producing longer strings. 

From general considerations at a second order phase transition 
we can expect that all characteristic lengths will present 
divergences as $T\rightarrow T_c$. 
$a(\beta)$ is no exception. Note however that while $\xi(\beta)$, 
the correlation length associated with the connected field 
2-point function, measures isotropic correlations of field 
excitations, $a$ refers to correlations along the strings. 
It is therefore interesting to measure the critical
exponents associated with $a$. Fig.~\ref{fig3}b shows the best fits
to $a(\beta)$ at criticality: 
\begin{eqnarray}
&& a(\beta) -1 \propto 1/(\beta-\beta_c)^{\nu_a^+}, \qquad \beta > \beta_c 
\label{e5}\\
&& a(\beta) -1 \propto 1/(\beta_c-\beta)^{\nu_a^-}, \qquad \beta < \beta_c, 
\nonumber
\end{eqnarray}
with $\nu_a^+=\nu_a^-=0.5 \pm 0.1$. These values are typical of mean-field 
predictions and differ from $\nu=0.671$ \cite{Zinn}, 
associated with $\xi$.

Above the critical point the behavior of A is given 
by $A(\beta) \simeq 0.092 + 0.006~\beta$. 
Note that in addition, given $\gamma$ and above $T_c$ 
($\sigma_{\rm eff}=0$), the total density of loops
\begin{eqnarray}
\rho_{\rm loop} = 
\sum_{l=4}^{\infty} l <n(l)> = {A \over 4^{(\gamma-2)} (\gamma-2)}
\quad (\gamma>2)
\end{eqnarray}
is proportional to $A$ in the infinite volume.
For $\beta=0$, when the distribution becomes Brownian 
the value of $A$ does indeed exactly coincide with our 
measurement of $\rho_{\rm loop} (0)=A(0)=0.083$ in \cite{Us}.    

\begin{figure}
\centerline{\psfig{file=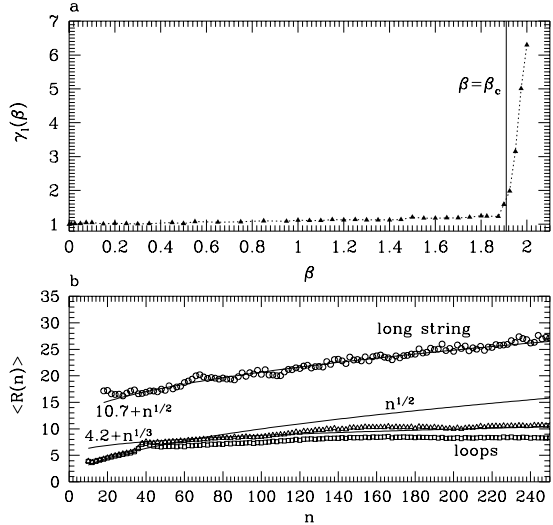,width=3in}}
\caption{a) $\gamma_l(\beta)$ for the 
long string length distribution and b) $R(n)$ at $\beta=1.8$
for loops ($\triangle$) and long strings ($\bigcirc$) and for 
loops at $\beta=2.05$ ($\Box$).
Solid curves show the behavior of free random ($\alpha=0.5$) 
and self-seeking ($\alpha=1/3$) walks. }
\label{fig4}
\end{figure}

In order to complete the picture of the role of vortex string 
fluctuations in the thermodynamics of the theory we characterize in 
Fig.~\ref{fig4}, the length distribution of long strings and show 
the mean distance between two string segments $R(n)$ 
as a function of the number of steps (in lattice units) along the string.
The distribution of long string, is notoriously
difficult to measure because the total number of long strings 
in any computational domain is typically only a few.
Here we could probe 600-1200 independent string networks, 
a large number compared to any study in a computational domain of this size.
However, in such a set of data the number of long strings of 
a given length is never more than one in the total ensemble 
of measurements and a direct fit to the data  becomes impossible. 
To obviate this problem we adopted  a 
maximum likelihood test,  see eg. \cite{Stats},  
to fit the data from all samples to a  distribution of the form 
$A l^{-\gamma_{l}}$. In the free random walk approximation 
$\gamma_{l}=1$, as the result of the periodicity of the 
boundary conditions \cite{ERA}.

Fig.\ref{fig4}a shows the temperature dependence of 
$\gamma_{l}$. 
For $T>T_c$, $\gamma_{l}$ is close to the Brownian 
value $\gamma_{l}=1$ converging to it as $\beta \rightarrow 0$, as 
$\gamma_l \simeq 1+ 0.1147~\beta$.
In the small band of temperatures below $T_c$ for which we identified
long string, $\gamma_{l}$ assumes much larger values. 

This sharp increase in $\gamma_{l}$ is an indication that the
distribution no longer has the simple form of a power-law. In 
analogy to the form of the loop distribution it marks the 
onset of exponential suppression. Note that it was necessary to look 
at strings classified as long to notice this suppression, allowing us
a more precise insight into the temperature variation of $\sigma_{\rm eff}$,
just below criticality. 
This behavior reinforces our previous conjecture that true long string 
is present in the system only above $T_c$ which would, 
in the infinite volume limit, coincide with $T_H$. 
  
A complementary test on the interacting nature of the strings 
is provided by $R(n)$, shown in Fig.~\ref{fig4}b. 
Generally one expects $R(n) \sim n^\alpha$,
where $n$ is the number of steps in lattice units. For Brownian
walks it is well known that $\alpha=0.5$. On the other hand 
$\alpha>0.5$ if the interactions are repulsive, 
($\alpha \simeq 0.59$ for self-avoiding walks)
and $\alpha\simeq 1/3$ if they are attractive at low $T$ \cite{Binder}.  

In $U(1)$ scalar field theory two string segments, at distance $d$ 
interact with potential $V(d) \simeq {\vec  n_1}.{\vec n_2} \ln(r/m)$, 
for  $d \geq m^{-1}$, the width of the string. 
Here $\vec n_i$ is the unit vector normal to the $i^{th}$ string segment 
times the topological charge of the vortex. 
Thus vortices interact with a potential which grows 
with distance. At short distances $d<m^{-1}$ the interactions vanish
since the superposition of two vortices 
(or of a vortex-antivortex) is also a solution of the static 
field equations. 

Small loops of string are the direct equivalent in 3D 
of vortex-antivortex pairs in 2D, since each segment 
will tend on the average to see, normal to the string, another
string segment with opposite vorticity. This fact permits the 
nucleation of small loops of string below $T_c$.

As $T$ increases longer loops can be nucleated.
The nature of the interactions  makes it advantageous
to create loops where the correlations between segments of opposite
vorticity are maximal. Consequently loops behave on the average 
as random walks for a short distance, while 
the interactions are irrelevant, and beyond become self-seeking, 
see Fig.~\ref{fig4}b. This distance increases with $T$. 

As the critical point is approached  
the total string density mounts to a universal critical  
value $\rho_{\rm tot} (\beta_c) \simeq 0.2$. 
At this point enough string exists 
in the system that the effects of the interactions shield each 
others out efficiently. Precisely then a population of long strings can be 
nucleated, interacting with itself through the background loop sea 
with a potential that is almost vanishing. In this sense long strings are
born quasi-Brownian. This fact is demonstrated in Fig.~\ref{fig4}b. After 
a short transient for small distances, where they are 
self-avoiding (otherwise they would have become a loop), 
long strings assume an extraordinarily clear random walk 
behavior, corresponding to Hausdorf dimension $D_H\simeq 2$. 
This establishes the difference between the two string populations, 
since loops remain self-seeking ($D_H \simeq 3$) 
even for $T>T_c$, see Fig.~\ref{fig4}.

In conclusion, we have found quantitative evidence that the phase transition
in 3D scalar $U(1)$ field theory proceeds by vortex string nucleation and 
in particular is triggered by long string proliferation, 
as often suggested \cite{PT}.
We established that a population of quasi-Brownian long strings
exists above $T_c$ only, and that it is different 
from small loops which remain self-seeking. 
We characterized the length distribution of both loops and long strings,
and measured associated universal critical exponents. 
The latter provide the first direct constraints on field theories dual
to the $U(1)$ scalar model in 3D. The critical
behavior of the length distributions confirms our previous claim 
that the Hagedorn phase transition in the strings coincides 
with the critical point of the theory. 
The Ginzburg temperature, ($\beta_G\simeq 2.34$), 
seems to play no particular role in the thermodynamics.

We thank M.~Hindmarsh, J.~Kottmann and A. Schackel for useful discussions,
the Department of Computer Science at the T.~U. 
of Berlin for usage of their Cray T3E  and A.~Yates for his string 
tracing routine. This work was partially supported by the 
European Science Foundation.
NDA is supported by JNICT, contract PRAXIS XXI/BD/13765/97.

\end{document}